\PassOptionsToPackage{unicode}{hyperref}
\PassOptionsToPackage{hyphens}{url}
\PassOptionsToPackage{dvipsnames,svgnames,x11names}{xcolor}
\documentclass[
  letterpaper,
  DIV=11,
  numbers=noendperiod,
  letterpaper,
  man,
  natbib,
  11pt]{scrartcl}

\usepackage{amsmath,amssymb}
\usepackage{iftex}
\ifPDFTeX
  \usepackage[T1]{fontenc}
  \usepackage[utf8]{inputenc}
  \usepackage{textcomp} 
\else 
  \usepackage{unicode-math}
  \defaultfontfeatures{Scale=MatchLowercase}
  \defaultfontfeatures[\rmfamily]{Ligatures=TeX,Scale=1}
\fi
\usepackage{lmodern}
\ifPDFTeX\else  
\fi
\IfFileExists{upquote.sty}{\usepackage{upquote}}{}
\IfFileExists{microtype.sty}{
  \usepackage[]{microtype}
  \UseMicrotypeSet[protrusion]{basicmath} 
}{}
\makeatletter
\@ifundefined{KOMAClassName}{
  \IfFileExists{parskip.sty}{%
    \usepackage{parskip}
  }{
    \setlength{\parindent}{0pt}
    \setlength{\parskip}{6pt plus 2pt minus 1pt}}
}{
  \KOMAoptions{parskip=half}}
\makeatother
\usepackage{xcolor}
\makeatletter
\ifx\paragraph\undefined\else
  \let\oldparagraph\paragraph
  \renewcommand{\paragraph}{
    \@ifstar
      \xxxParagraphStar
      \xxxParagraphNoStar
  }
  \newcommand{\xxxParagraphStar}[1]{\oldparagraph*{#1}\mbox{}}
  \newcommand{\xxxParagraphNoStar}[1]{\oldparagraph{#1}\mbox{}}
\fi
\ifx\subparagraph\undefined\else
  \let\oldsubparagraph\subparagraph
  \renewcommand{\subparagraph}{
    \@ifstar
      \xxxSubParagraphStar
      \xxxSubParagraphNoStar
  }
  \newcommand{\xxxSubParagraphStar}[1]{\oldsubparagraph*{#1}\mbox{}}
  \newcommand{\xxxSubParagraphNoStar}[1]{\oldsubparagraph{#1}\mbox{}}
\fi
\makeatother

\providecommand{\tightlist}{%
  \setlength{\itemsep}{0pt}\setlength{\parskip}{0pt}}\usepackage{longtable,booktabs,array}
\usepackage{calc} 
\usepackage{etoolbox}
\makeatletter
\patchcmd\longtable{\par}{\if@noskipsec\mbox{}\fi\par}{}{}
\makeatother
\IfFileExists{footnotehyper.sty}{\usepackage{footnotehyper}}{\usepackage{footnote}}
\makesavenoteenv{longtable}
\usepackage{graphicx}
\makeatletter
\newsavebox\pandoc@box
\newcommand*\pandocbounded[1]{
  \sbox\pandoc@box{#1}%
  \Gscale@div\@tempa{\textheight}{\dimexpr\ht\pandoc@box+\dp\pandoc@box\relax}%
  \Gscale@div\@tempb{\linewidth}{\wd\pandoc@box}%
  \ifdim\@tempb\p@<\@tempa\p@\let\@tempa\@tempb\fi
  \ifdim\@tempa\p@<\p@\scalebox{\@tempa}{\usebox\pandoc@box}%
  \else\usebox{\pandoc@box}%
  \fi%
}
\def\fps@figure{htbp}
\makeatother
\NewDocumentCommand\citeproctext{}{}

\makeatletter
 \let\@cite@ofmt\@firstofone
 \def\@biblabel#1{}
 \def\@cite#1#2{{#1\if@tempswa , #2\fi}}
\makeatother
\newlength{\cslhangindent}
\newlength{\csllabelwidth}
\newenvironment{CSLReferences}[2] 
 {\begin{list}{}{%
  \setlength{\itemindent}{0pt}
  \setlength{\leftmargin}{0pt}
  \setlength{\parsep}{0pt}
  \ifodd #1
   \setlength{\leftmargin}{\cslhangindent}
   \setlength{\itemindent}{-1\cslhangindent}
  \fi
  \setlength{\itemsep}{#2\baselineskip}}}
 {\end{list}}
\usepackage{calc}

\KOMAoption{captions}{tableheading}
\makeatletter
\@ifpackageloaded{caption}{}{\usepackage{caption}}
\AtBeginDocument{%
\ifdefined\contentsname
  \renewcommand*\contentsname{Table of contents}
\else
  \newcommand\contentsname{Table of contents}
\fi
\ifdefined\listfigurename
  \renewcommand*\listfigurename{List of Figures}
\else
  \newcommand\listfigurename{List of Figures}
\fi
\ifdefined\listtablename
  \renewcommand*\listtablename{List of Tables}
\else
  \newcommand\listtablename{List of Tables}
\fi
\ifdefined\figurename
  \renewcommand*\figurename{Figure}
\else
  \newcommand\figurename{Figure}
\fi
\ifdefined\tablename
  \renewcommand*\tablename{Table}
\else
  \newcommand\tablename{Table}
\fi
}
\@ifpackageloaded{float}{}{\usepackage{float}}
\floatstyle{ruled}
\@ifundefined{c@chapter}{\newfloat{codelisting}{h}{lop}}{\newfloat{codelisting}{h}{lop}[chapter]}
\floatname{codelisting}{Listing}

\makeatother
\makeatletter
\@ifpackageloaded{caption}{}{\usepackage{caption}}
\@ifpackageloaded{subcaption}{}{\usepackage{subcaption}}
\makeatother

\usepackage{bookmark}

\IfFileExists{xurl.sty}{\usepackage{xurl}}{} 
\hypersetup{
  pdftitle={The ultimate issue error in scientific inference: mistaking parameters for hypotheses},
  pdfauthor={Stanley E. Lazic\^{}\{1,*\}},
  colorlinks=true,
  linkcolor={blue},
  filecolor={Maroon},
  citecolor={Blue},
  urlcolor={Blue},
  pdfcreator={LaTeX via pandoc}}

\title{The ultimate issue error in scientific inference: mistaking
parameters for hypotheses}
\author{Stanley E. Lazic\(^{1,*}\)}
\date{}

\begin{document}
\maketitle

\begin{center}
1. Prioris.ai Inc., 459-207 Bank St., Ottawa ON K2P 2N2, Canada

$^{*}$Corresponding author: stan.lazic@cantab.net
\end{center}

\section{Abstract}\label{abstract}

Statistical inference often conflates the probability of a parameter
with the probability of a hypothesis, a critical misunderstanding termed
the \emph{ultimate issue error}. This error is pervasive across the
social, biological, and medical sciences, where null hypothesis
significance testing (NHST) is mistakenly understood to be testing
hypotheses rather than evaluating parameter estimates. Here, we advocate
for using the Weight of Evidence (WoE) approach, which integrates
quantitative data with qualitative background information for more
accurate and transparent inference. Through a detailed example involving
the relationship between vitamin D (25-hydroxy vitamin D) levels and
COVID-19 risk, we demonstrate how WoE quantifies support for hypotheses
while accounting for study design biases, power, and confounding
factors. These findings emphasise the necessity of combining statistical
metrics with contextual evaluation. This offers a structured framework
to enhance reproducibility, reduce false interpretations, and foster
robust scientific conclusions across disciplines.

\newpage{}

\section{Introduction}\label{introduction}

Suppose fingerprints are found at a crime scene and police have detained
a suspect. The police hypothesise that if the suspect is guilty, their
fingerprints should match those found at the scene. Upon testing, the
forensic team concludes that there is a one in a million chance that the
fingerprints originated from someone other than the suspect. What is the
probability that the suspect is guilty? This is not a trick question
about conditional probabilities, p-value interpretations, or the
prosecutor's fallacy. The answer is: we cannot determine the probability
of guilt given the current information. If the crime happened in the
suspect's home, their fingerprints would be found everywhere. Therefore,
finding their fingerprints at the crime scene does not imply guilt
because \(P(\text{fingerprint} | \text{not guilty}) \approx 1\). If,
however, the suspect's fingerprints are found in the home of someone
they do not know, the fingerprint evidence may be highly suggestive of
guilt.

This example highlights two key points. First, qualitative background
information is critical for determining what a piece of evidence says
about a hypothesis. Second, the degree to which the evidence implies
guilt has little to do with the probability of a match; these are
probabilities for different events. The probability of a fingerprint
match informs the probability of guilt, but assuming that the
probability of a match \emph{equals} the probability of guilt is known
as the ultimate issue error (Aitken et al. 2021). The ``ultimate issue''
is whether the suspect is guilty, and the error arises from substituting
another probability; the 1:1,000,000 match probability.

A hypothesis is a testable statement or proposition. It is often
expressed as a prediction based on some theory, model, or background
information. A parameter is a quantitative component of a statistical
model, which often represents some characteristic of a population. An
example of a hypothesis is: ``If this drug is effective, blood pressure
in the drug group will be lower than blood pressure in the control
group''. The parameter might be the mean difference in blood pressure
between the two groups, denoted by \(\delta\). If \(\delta < 0\), all we
can conclude is that the hypothesis is now more \emph{plausible} than
before obtaining the data, but not by how much (Polya 1954). In other
words, \(P(\delta < 0) \neq P(\text{Drug is effective})\). Observing
that \(\delta < 0\) is a necessary but not sufficient condition for
concluding that the drug is effective.

Unfortunately, the ultimate issue error is common in the social,
biological, and medical sciences when testing hypotheses. Here, the
ultimate issue is the probability that a hypothesis is true, but the
probability calculated is whether a parameter in a statistical model
equals a given value. We refer to this procedure as null hypothesis
significance testing (NHST) when it is really null parameter
significance testing (NPST).\footnote{Further complicating matters,
  frequentist hypothesis testing does not directly test a hypothesis;
  instead, it tests if the data are inconsistent with the null
  hypothesis.} The relationship between parameter values and quantities
derived from them, such as p-values or Bayes factors, have no direct
quantitative relationship with scientific hypotheses.

We argue below that parameters and hypotheses are distinct entities and
that parameter testing is usually quantitative, whereas hypothesis
testing is mostly qualitative and subjective. Next, the problem of
confusing the two will be described. Finally, an approach to quantify
the support for a hypothesis will be presented.

\subsection{\texorpdfstring{\(P(\text{Parameter}) \neq P(\text{Hypothesis})\)}{P(\textbackslash text\{Parameter\}) \textbackslash neq P(\textbackslash text\{Hypothesis\})}}\label{ptextparameter-neq-ptexthypothesis}

This section describes three reasons why the probability of a parameter
usually does not equal the probability of a hypothesis. First,
qualitative background information is always needed to interpret a
parameter's meaning in light of a hypothesis. Consider the following
example. Suppose we are testing if Drug Z is effective for treating
depression. We run a study with 200 subjects and find that the drug
group has a 22\% improvement compared to the control group, with
\(p = 0.004\) (or a Bayes Factor of 16.7, or a posterior probability of
0.998 if you prefer a Bayesian analysis). Assume that a 22\% improvement
is clinically relevant. What is the probability that Drug Z is effective
for treating depression?

A typical concluding statement in a manuscript would be ``the data
support the effectiveness of Drug Z for depression (\(p = 0.004\)),''
where the p-value is supposed to justify the preceding sentence, but it
fails to do so. Consider the potential background information about this
study and how it influences the probability of the drug's effectiveness:

\begin{itemize}
\tightlist
\item
  Double-blinded versus open-label/unblinded study.
\item
  Randomised controlled trial versus observational study.
\item
  Hard primary end point (suicide) versus soft primary end point
  (self-reported depression).
\item
  A plausible mechanism of action based on preclinical data versus no
  known mechanism of action and no preclinical data.
\item
  All subjects completed the study versus twice as many subjects in the
  drug group were lost to follow-up.
\item
  Published as a registered report versus a non-registered report.
\item
  Funded and run by an independent academic group versus the company
  that developed the drug.
\item
  None of the authors had previous papers retracted versus the senior
  author had several papers retracted (he blames his postdoc).
\item
  Published in the New England Journal of Medicine versus a predatory
  journal.
\item
  It's the second study to obtain a positive result with this drug
  versus it's the first study.
\end{itemize}

Many of these points fall under the familiar categories of internal and
external validity, risk of bias, and reproducibility. Few would be
convinced of the drug's effectiveness if this was an unblinded
observational study using a soft endpoint with massive drop-out in the
drug group, run by a lead investigator of dubious reputation who works
for the company that produces the drug, and published a few days after
initial submission in a predatory journal. It does not matter how small
the reported p-value or how large the Bayes factor is, there are too
many problems with the study to consider the results credible. Similar
to jurors weighing all the evidence to reach a verdict, researchers will
take the numeric results and weigh them with relevant background
information to determine if the results are convincing.

For both jurors and researchers, this is a subjective process, and is no
different from what every researcher informally does when reading a
paper (``this is solid'' or ``I doubt this will replicate''). Some may
think that subjective judgements have no place in science, but they are
unavoidable, and they are already performed when assessing the risk of
bias in clinical studies (Higgins et al. 2011). Note how the criteria
can combine to influence the overall judgement. A blinded study with a
soft endpoint may be convincing, as might a study without blinding but
with a hard endpoint. However, an unblinded study with a soft endpoint
might have a risk of bias too high to be convincing.

The second reason why the probability of a parameter usually does not
equal the probability of a hypothesis is because there is often a
one-to-many relationship between hypotheses and parameters; studies
often have multiple outcomes that are associated with multiple
parameters. For example, a study may assess depression with the Hamilton
Depression Ration Scale, the Montgomery-Asberg Depression Rating Scale,
and a simple self-report measure, all of which will be associated with a
parameter. These parameters will rarely be equally distant from the null
(on the appropriately standardised scale), and may even conflict.
Arbitrarily defining a primary outcome to test the hypothesis only
side-steps the problem. If there is a need to measure multiple outcomes
to test a hypothesis, it is likely because the true outcome of interest
is not directly observable or is ill-defined. This raises issues of
construct validity (how do we know we are actually measuring
depression?) and content validity (how do we know we're measuring all
aspects of it?), further weakening the link between a parameter based on
a measured outcome and a hypothesis.

Finally, parameters critically depend on the details of the experiment
or study as well as the statistical model. For example, the drug might
work better in patients with severe depression than patients with mild
depression. The parameter will therefore differ based on the study's
inclusion and exclusion criteria. Even worse is when the parameter
depends on unknown or unmeasured population characteristics. For
example, suppose the drug works better in patients with predominately
affective symptoms (feelings of sadness or hopelessness) versus patients
with predominately physiological symptoms (tiredness, lack of energy,
sleep disturbances), but this is unknown to the researcher. The effect
(parameter) will then largely reflect the proportion of these patient
subtypes in the experiment.

The value of a parameter and its standard error (SE) depend on the
statistical model in which they are embedded. Models may vary in their
likelihood function (e.g.~normal or Poisson) the predictor variables
included, whether nonlinearities or interactions are included, and so
on. Several models may be reasonable, and the parameter and/or its SE
will depend on the selected model. Therefore, no unique mapping between
a parameter and a hypothesis exists. Even for simple experiments such as
our drug example, whether the drug is effective could be defined as the
mean difference between groups, the median difference, the ratio of
means or medians, or something else.

The probability of a parameter only approximates the probability of a
hypothesis when the hypothesis is a precise numeric value, which is more
common in physics and engineering. For example, based on his theory of
relativity, Einstein predicted that the sun would deflect star light by
1.75 seconds of arc (Earman and Glymour 1980). This precisely
hypothesised value was then compared with experimental observations and
found to be in accordance with them. However, even in this case,
background information was important. Experimental observations that
failed to support Einstein's prediction would only count against his
theory if the measuring instruments were appropriate for the task, were
calibrated and working properly on the day, and so on. Few if any
experiments in the biological and social sciences make precise numeric
predictions, and rarely even make a directional prediction, as evidenced
by the extensive use of two-tailed tests.

\subsection{Why make the hypotheses/parameter
distinction?}\label{why-make-the-hypothesesparameter-distinction}

Failure to distinguish between these probabilities can lead one to
believe that a hypothesis has much more support than is justified. This
misinterpretation can also spread to the wider public, including
journalists and policymakers, where the consequences can be severe.

In scientific publications, researchers must convince skeptical
colleagues of the validity of their claims. The distinction between the
probability of parameters and hypotheses emphasises the fact that small
p-values alone do not provide convincing evidence. Researchers must also
use appropriate methods, rule out competing explanations, ensure that
the results are unbiased and therefore credible, ensure the construct is
appropriately operationalised, and so on. Focusing on aspects that
increase the probability of a hypothesis will lead to better
experiments, compared to exclusively focusing on aspects that lead to a
higher probability of rejecting a null parameter value, such as
increased sample size.

\subsection{Parameters informing
hypotheses}\label{parameters-informing-hypotheses}

No method has been generally adopted for moving from a quantitative
value for a parameter to a quantitative statement about a proposition,
such as a scientific hypothesis. However, Pierce described an approach
in 1878 (Peirce 2014), which has been modified and applied by many
others (Good 1950; Jaynes 2003; Edwards 1992; Pardo and Allen 2007;
Allen and Pardo 2019; Fairfield and Charman 2022). Essentially, it is a
comparison of the relative likelihood or plausibility of the evidence
(\(E\)) under two competing hypotheses or explanations (\(H_1\) and
\(H_0\)), expressed as a ratio of one hypothesis to another, giving a
likelihood ratio (LR).

\begin{equation}\phantomsection\label{eq-LR}{
LR = \dfrac{P(E|H_1)}{P(E|H_0)}.
}\end{equation}

Continuing our legal example, let \(E\) represent the fingerprint
evidence, \(H_1\) represent the prosecution's hypothesis that the
defendant is guilty, and \(H_0\) represent the defence's position that
the accused is not guilty and the fingerprint match is the result of
chance. If the accused is guilty, we expect to find their fingerprints
at the crime scene, so \(P(E|H_1) \approx 1\). If they are not guilty,
the random match probability is \(P(E|H_0) = 1/1,000,000\), which gives

\[
LR = \dfrac{P(E|H_1)}{P(E|H_0)} = \dfrac{1}{10^{-6}} = 10^6,
\]

\noindent interpreted as the evidence being one million times more
likely under the prosecution's position. However, this is not the
likelihood that the accused is guilty, only the likelihood that they are
the source of the fingerprint. To establish guilt, this evidence must be
considered with all the other evidence, such as a lack of motive and a
strong alibi, for example. Reasonable people would agree that having no
motive and an alibi decreases the probability of guilt, but opinion will
differ as to how much. This is the crux of the problem, and why criminal
cases ask twelve jurors for their judgement.

The LR quantifies how strongly evidence supports one hypothesis or the
other, but Equation~\ref{eq-LR} does not consider the other evidence
required to interpret the match probability. Let's augment
Equation~\ref{eq-LR} to include all the evidence (\(E^*\)) relevant for
assessing how the match probability influences the probability of guilt,
as well as all the background information (\(I\)), which includes
common-sense information such as ``having a motive increases the
likelihood of guilt''

\begin{equation}\phantomsection\label{eq-LR-star}{
LR^* = \dfrac{P(E^*|H_1, I)}{P(E^*|H_0, I)}.
}\end{equation}

The augmented likelihood ratio (\(LR^*\)) combines the quantitative
random match probability defined in Equation~\ref{eq-LR} with
qualitative information about motive, alibi, the probability of a
laboratory error, the uncertainty about the reference population used to
calculate the 1:1,000,000 number, and so on. At this point, judgement is
required. It is the responsibility of the judge or jury in a criminal
trial, or the editor, peer-reviewer, or other scientist when evaluating
scientific findings.

Rather than dealing with \(LR^*\), it is often more convenient to take
the logarithm. By convention, a base-10 logarithm is used and the
resulting log-likelihood ratio is multiplied by 10 to give decibel
units: \(10\, \text{Log}_{10}(LR^*)\). This quantity is known as the
weight of evidence (WoE) (Peirce 2014; Good 1950). Logarithms have
several advantages. First, independent pieces of evidence can be added
together to get an overall WoE. Second, a logarithmic decibel scale is
more intuitive (once you are familiar with it). For example, WoE = 0
means that both hypotheses have equal support, and WoE \textgreater{} 0
favours \(H_1\) while WoE \textless{} 0 favours \(H_0\). The WoE is
symmetric whereas the LR is not. For example, a LR of 15 favours \(H_1\)
to the same degree that a LR of 0.07 favours \(H_0\), but this is not
obvious (\(0.07 = 1/15\)). The corresponding values on a logarithmic
scale are \(10\, \text{log}_{10}(15) = 11.8\) and
\(10\, \text{log}_{10}(1/15) = -11.8\), making this symmetric
relationship clear. Furthermore, the difference between 0.91 and 0.99
seems larger than that between 0.99 and 0.999, yet both are 10 decibels
apart. Finally, psychophysics has shown that perceptions of stimulus
intensity are often proportional to the logarithm of intensity (Colman
2009), and by analogy, the degree to which beliefs should be updated is
proportional to the logarithm of evidence. There is also recent evidence
that people can more easily combine information using log ratios,
compared with ratios or probabilities (Juslin et al. 2011).

Table 1 shows the relationship between the WoE, odds, and probability of
hypothesis \(H_1\) over \(H_0\). A 10 unit increase in the WoE
corresponds to a \(10\times\) increase in the odds. Three decibels makes
\(H_1\) about twice as likely as \(H_0\), and 12 decibels corresponds to
about a 0.95 probability in favour of \(H_1\) over \(H_0\).

\begin{longtable}[]{@{}lll@{}}
\caption{Relationship between the weight of evidence (WoE) for a
hypothesis (\(H\)) provided by the evidence (\(E\)), odds, and
probabilities.}\tabularnewline
\toprule\noalign{}
WoE(\(H : E\)) & Odds & Probability \\
\midrule\noalign{}
\endfirsthead
\toprule\noalign{}
WoE(\(H : E\)) & Odds & Probability \\
\midrule\noalign{}
\endhead
\bottomrule\noalign{}
\endlastfoot
0 & 1:1 & 0.5 \\
3 & 2:1 & 0.67 \\
6 & 4:1 & 0.8 \\
10 & 10:1 & 0.91 \\
12 & 19:1 & 0.95 \\
20 & 100:1 & 0.99 \\
30 & 1000:1 & 0.999 \\
\end{longtable}

\(P(E^*|H)\) in Equation~\ref{eq-LR-star} is the probability of the
evidence given a hypothesis, but we are interested in the probability of
a hypothesis given the evidence: \(P(H|E^*)\). Bayes' Theorem is the
standard way to convert from the first probability to the second, but it
requires one additional component: the prior probability of each
hypothesis \(P(H,I)\). In a legal setting, this term captures the
presumption of innocence. Combining all of these terms gives
\(\text{WoE}(H_1 : E^*,I)\), which is read as ``the weight provided by
\(E^*\) and \(I\) for hypothesis \(H_1\)''

\begin{equation}\phantomsection\label{eq-WoE}{
\text{WoE}(H_1 : E^*,I) = 10\, \text{log}_{10} \left[ \dfrac{P(E^*|H_1, I)}{P(E^*|H_0, I)} \right] + 10\, \text{log}_{10} \left[ \dfrac{P(H_1, I)}{P(H_0,I)} \right].
}\end{equation}

The above procedure maps cleanly from a legal setting to a scientific
one: \(E\) is the evidence for an effect or association, and for
simplicity we can take \(p < 0.05\) as evidence for an
effect\footnote{We interpret a small p-value as evidence that the
  experiment demonstrated an effect; we are merely using this as an
  indicator that ``something happened''. Any criterion can be used such
  as Bayes Factors or posterior densities past some threshold.}. \(E^*\)
includes the details of the study's design, data collection, analysis,
and potential biases as well as any conflicts of interest. \(H_0\) and
\(H_1\) are the null and alternative hypotheses, respectively. \(I\)
once again is all our background information, such as ``unblinded
studies are more likely to be biased''.

How can we quantify these terms? \(P(E^*|H_0, I)\) is the probability of
observing a significant result if the null hypothesis is true, which is
the probability of a false positive. This quantity is usually set to
\(\alpha = 0.05\) by the experimenter. However, the actual probability
of a false positive can be larger than 0.05 because of biases, perverse
incentives, questionable research practices (QRP), incompetence, or
fraud. A judgement regarding how much \(\alpha\) should be increased
from the nominal 0.05 value can be made by anyone evaluating the
experiment. If available, estimates of fraud rates or QRPs in a field
can be informative, as will the specific details of the study design and
execution. Note that the probability of a false positive result is not
the calculated p-value -- it is based on \(\alpha\), and adjusted as
needed based on the other evidence.

\(P(E^*|H_1, I)\) is the probability of observing a significant result
when the alternative hypothesis is true, and therefore represents the
statistical power of the experiment. This value might be reported if a
power calculation was performed. Or, it might be possible to estimate
the power from past studies if no information regarding the power is
provided. For example, Button et al. (2013) reported that the average
power in neuroscience experiments was around 21\%. The reported or
estimated power provides a starting value, but it will also need to be
updated to reflect the details of the experiment. Note that we are not
referring to retrospective or observed power, which is calculated from
the data and has a 1:1 mapping with the p-value (Hoenig and Heisey 2001;
Levine and Ensom 2001; Senn 2002).

Power and sample size calculations are often simple approximations of
the experiment to be conducted; they may use unrealistic effect sizes or
within-group variances, and may exclude relevant aspects of the design.
The actual power of the hypothetical antidepressant experiment may be
lower than the calculated value due to the following reasons:

\begin{itemize}
\tightlist
\item
  The dose was too low, so the effect would be small.
\item
  The drug was given for too short a duration before assessing its
  effectiveness.
\item
  The drug was given to some subjects who were unlikely to respond or
  benefit (e.g.~those with severe depression). The overall effect will
  therefore be diluted.
\item
  The dose-response relationship is not linear (e.g.~inverted ``U''
  shape), and the drug was given at too high a dose.
\item
  A sub-optimal route of administration or formulation was used. For
  example, the best results are via injection, but it was given orally
  for logistical reasons.
\item
  The subjects included in the study have mild symptoms, and there is
  little room for the drug to demonstrate an improvement (floor effect).
\item
  A surrogate outcome for depression is used that has a weak
  relationship with the true outcome of interest.
\item
  The primary outcome had many missing values, reducing the expected
  sample size.
\end{itemize}

Although most people will agree that the above points reduce the power
of an experiment, there will be disagreement as to how much. Judgement,
once again, is required to determine a suitable value for the actual
power of the experiment.

The final component of Equation~\ref{eq-WoE} is the prior, which is the
term to the right of the addition sign. The ratio represents the
relative likelihood of the two hypotheses, before considering the data
from the present experiment. This value could be based on results from
past experiments, or a general assessment of the plausibility of each
hypothesis. If this is the first experiment, or if we want to draw
conclusions independent of previous experiments, then giving the two
hypotheses equal prior plausibility will make the log of the ratio equal
to zero, thus dropping this term from consideration. As a result, the
WoE will only be influenced by the log-likelihood ratio.

Another way to think of the LR is in the context of a diagnostic test,
where \(H_1\) = ``person has the disease'', \(H_0\) = ``person does not
have the disease'', and \(E\) is a positive test result, giving

\begin{equation}\phantomsection\label{eq-PLR}{
\dfrac{P(E|H_1)}{P(E|H_0)} = \dfrac{P(\text{positive test | disease})}{P(\text{positive test | no disease})} = \dfrac{\text{sensitivity}}{1 - \text{specificity}} = \dfrac{\text{sensitivity}}{\text{false positive rate}}.
}\end{equation}

The numerator is also known as the sensitivity (the proportion of
positive tests among all the people with the disease), and the
denominator is equal to one minus the specificity (the proportion of
people who test negative among those who do not have the disease). The
sensitivity and specificity are operating characteristics of a given
diagnostic test. Similarly, an experiment may be characterized as having
a certain sensitivity for detecting an effect (power) and the ability to
correctly detect the absence of an effect (expressed as the false
positive rate, or 1 - true negative rate).

Suppose, for example, that the antidepressant trial had a significant
difference between groups (i.e.~a positive test). The study was powered
for 80\%, but the drug was administered at too low a dose. Consequently,
the probability of observing an effect is reduced, say the power is now
only 60\%. Furthermore, suppose that the researchers used
\(\alpha = 0.05\) as their significance threshold. The informed consent
form indicated that nausea was a likely side-effect of the drug, and
80\% of patients on the drug indicated feeling nauseated at some point
(compared with 5\% of the control group). Hence, patients experiencing
nausea may have concluded that they received the drug, and are now
effectively unblinded. Given the subjective nature of the primary
outcome, patients may expect to improve, and this expectation may bias
the results such that the probability of a false positive is now
increased from the nominal value of 0.05 to, say, 0.15. In the context
of diagnostic tests, priors correspond to the base-rate or the
prevalence of the disease in the relevant population. For this example,
results from previous studies are not used. The prior ratio therefore
equals one, and the log of the prior ratio equals zero and does not
contribute to the result. Plugging these numbers into
Equation~\ref{eq-WoE} gives

\[
\text{WoE}(H_1 : E^*,I) = 10\, \text{log}_{10} \left[ \dfrac{0.6}{0.15} \right] + 10\, \text{log}_{10} \left[ 1 \right] = 6.02 + 0 \approx 6.
\]

Hence, despite a significant p-value, the WoE equals 6 (or 0.8
probability) indicating that \(H_1\) only has modest support relative to
\(H_0\) when all the evidence about the experiment is included. This
once again highlights the difference between a parameter being unlikely
given a hypothesis and the probability of the hypothesis -- the ultimate
issue.

If the trial returned a negative result (\(p>0.05\)), how likely is it
that the drug is \emph{ineffective}? We can perform a similar analysis,
but now let \(E\) = ``negative result''. Using the formula for a
diagnostic test we get

\begin{equation}\phantomsection\label{eq-NLR}{
\dfrac{P(E|H_1)}{P(E|H_0)} = \dfrac{P(\text{negative test | disease})}{P(\text{negative test | no disease})} = \dfrac{1 - \text{sensitivity}}{\text{specificity}} =  \dfrac{\text{false negative rate}}{\text{specificity}},
}\end{equation}

\noindent which gives

\[
\text{WoE}(H_1 : E^*,I) = 10\, \text{log}_{10} \left[ \dfrac{0.4}{0.85} \right] + 10\, \text{log}_{10} \left[ 1 \right] = -3.27 + 0 \approx -3.
\]

The WoE for \(H_1\) is -3, or equivalently, the WoE for \(H_0\) is 3,
which is negligible. We initially assumed that the hypotheses were
equally likely. After the study we have only a probability of
\(H_0 = 0.67\). Hence, the results of this experiment are uninformative
regardless of whether they are positive or negative. A positive result
only provides a WoE of 6 for \(H_1\) and a negative result gives a WoE
of 3 for \(H_0\). From these equations we can also see that if the
probability of a false positive is greater than the power, a study can
never provide evidence for \(H_1\).

\subsection{Real world example: COVID-19 and vitamin
D}\label{real-world-example-covid-19-and-vitamin-d}

Using United Kingdom (UK) Biobank data, Hastie et al. (2020) examined
the relationship between blood vitamin D levels (25-hydroxy vitamin D),
and the risk of COVID-19 infection. Complete data were available for
348,598 participants. In this data set, 2,724 COVID-19 tests were
conducted on 1,474 individuals, and 449 had a positive COVID-19 test.
This group was compared to the 1,025 people who tested negative, and the
347,124 people who were not tested. The outcome was confirmed COVID-19
infection, defined as at least one positive test result.

The main finding was that Vitamin D was associated with COVID-19
infection in a univariate logistic regression model (odds ratio = 0.99;
95\% CI 0.99 to 0.999; p = 0.013), but not after adjustment for
confounders (OR = 1.00; 95\% CI = 0.998 to 1.01; p = 0.208). The
author's conclusion was ``Our findings do not support a potential link
between vitamin D concentrations and risk of COVID-19
infection\ldots{}''. This paper has been cited nearly 600 times and
referred to as ``credible evidence'' by factcheck.org that low vitamin D
is not associated with a higher risk of COVID-19 infection.\footnote{https://www.factcheck.org/2020/06/does-vitamin-d-protect-against-covid-19/}

Using the above approach (e.g. Equation~\ref{eq-NLR}), we can assess
whether this study provides strong evidence for a lack of an association
between vitamin D levels and COVID-19 infection and whether the
conclusion is justified. Since we are only interested in quantifying the
evidence in this study, prior information will not be included. This
means the two hypotheses will be given equal initial probabilities of
0.5. The basic equation is then

\[
\dfrac{P(E|H_1)}{P(E|H_0)} = \dfrac{P(\text{negative result | association exists })}{P(\text{negative result | no association})}.
\]

We can initially take the denominator to be
\(1 - \alpha = 1 - 0.05 = 0.95\) (although this was not explicitly
stated by the researchers). However, the probability of a false positive
is slightly higher because the COVID-19 positive and negative groups
differed on almost all baseline variables (Table 1 in Hastie et al.
(2020)). Although these were adjusted for in the analysis, there may be
residual confounding due to nonlinear relationships or interactions
between variables. In addition, several analyses were conducted using
univariate or multivariate models and treating vitamin D levels as
continuous or binary. Let us modestly increase the probability of a
false positive from 0.05 to 0.1, and hence the denominator will be
\(1 - 0.1 = 0.9\)

The numerator is of greater interest because the study was negative, and
we want to know the probability of a negative result if in fact there is
an association. An initial value for the numerator is 1 - power. Hastie
et al. (2020) did not provide a power calculation, but we can get an
approximate estimate of the power based on the sample size (N=348,598),
the total number of positive cases (449), and the statistical model
(logistic regression). For the power calculation, assume that vitamin D
level is the only predictor and that it is a binary high/low value. The
outcome is testing positive or negative for COVID-19. We need to choose
a clinically meaningful effect size, and we can use an odds ratio of
0.80, which corresponds to a difference of 50 cases between the high and
low vitamin D groups. The power for this study is only 65\%\footnote{R
  code for the calculation is provided in the supplementary material.},
and therefore the numerator is \(1 - 0.65 = 0.35\). Entering these
numbers into our equation gives a value of

\[
\text{WoE}(H_1 : E^*,I) = 10\, \text{log}_{10} \left[ \dfrac{0.35}{0.90} \right] + 10\, \text{log}_{10} \left[ 1 \right] = -4.1.
\]

In other words, if we started with both hypotheses having an equal
probability of 0.5, after this study, the probability of no association
(\(H_0\)) has increased to only 0.72, which is a modest change and
hardly convincing. The situation is much worse however. This calculation
is based only on the study design and does not account for two critical
issues that considerably reduce the power.

First, vitamin D levels were not measured just before participants were
exposed to the virus, but 10-14 years earlier! The \emph{measured}
vitamin D levels are therefore a noisy proxy for the \emph{actual}
vitamin D levels. Jorde et al. (2010) estimated that vitamin D
measurements taken 14 years apart have a Pearson correlation of 0.42 to
0.52 (depending on the method used to adjust for seasonal variation),
and Meng et al. (2012) estimated a Spearman correlation of 0.61 when
measurements were 5.1 years apart. The additional noise introduced by
using a proxy variable will obscure any true association between vitamin
D levels and COVID-19 infection.

The second major issue is that the COVID-19 negative group were
comprised of 1,025 people who tested negative, and 347,124 people who
were not tested but just assumed to be negative. Many of these people
likely had COVID-19 but were not tested if they were not hospitalised,
and Davies, Mazess, and Benskin (2021) estimate that there could have
been nine times as many positive people in the ``negative'' group than
in the ``positive'' group. This introduces considerable
misclassification error, which will bias results towards the null
hypotheses of no association.

Taken together, these issues greatly reduce power from the already low
value of 65\%. Simulations could be conducted if an accurate power
estimate is important, but let's estimate the power to be 20\%, which
gives \(1 - 0.2 = 0.8\) for the numerator of the above formula, and
therefore \(\text{WoE}(H_1 : E^*,I) = -0.75\). In other words, the
probability of ``no association'' has moved from an initial value of 0.5
to a final value of 0.54! This study had additional issues that would
reduce power further (Davies, Mazess, and Benskin 2021), but let's stop
here. In summary, this study has provided \emph{zero evidence} for a
lack of an association between vitamin D and COVID-19 infection.

\section{Discussion}\label{discussion}

Researchers are not expected to calculate the WoE in order to supplement
the results of parameter testing. It is likely that researchers are
unaware of flaws or biases in their studies, otherwise they would have
designed them differently. Calculating the WoE is better done by
scientific colleagues, and possibly peer reviewers and editors.
Nevertheless, when designing a study, researchers may wish to consider
the WoE calculation in order to design a more effective experiment.

\subsection{Designing informative
experiments}\label{designing-informative-experiments}

From Equation~\ref{eq-PLR} and Equation~\ref{eq-NLR} we can see where to
focus to design informative experiments. For example, if a study is
initially designed with 80\% power and \(\alpha = 0.05\), is it better
to increase power or use a stricter alpha? Based on
Equation~\ref{eq-PLR}, we are better off decreasing the denominator
(false positives) of the LR than increasing the numerator (sensitivity).
The initial WoE (assuming no loss of power or bias) is
\(10\, \text{log}_{10}(0.8/0.05) \approx 12\). Increasing power to 95\%
only increases the WoE to
\(10\, \text{log}_{10}(0.95/0.05) \approx 12.8\), whereas decreasing
false positives to \(\alpha = 0.01\) increases the WoE to
\(10\, \text{log}_{10}(0.95/0.05) \approx 19\).

Therefore, to maximize the WoE an experiment can provide, researchers
should strive to keep the nominal false positive rate below
\(\alpha = 0.05\). This should come as no surprise, since minimising
bias, removing confounding, and ruling out alternative explanations are
effective ways to improve an experiment. However, when providing
evidence for the null hypothesis, it is better to minimise false
negatives (numerator of Equation~\ref{eq-NLR}), which can be verified
with similar calculations.

\subsection{Evaluating manuscripts and published
evidence}\label{evaluating-manuscripts-and-published-evidence}

The proposed approach relies on subjective judgement to weigh evidence
and interpret results, and interpretations may differ among researchers.
However, this is only a formalization of the process most scientists use
when reading a manuscript. Formalisation can help identify where the
problem(s) are, and if one is willing to put numbers to actual power and
the probability of a false positive, a quantitative estimate of the
support for one hypothesis over another can be derived. This process is
no different from what peer reviewers and editors informally do when
deciding to accept or reject a manuscript based on the quality of the
research. It is an empirical question whether such an approach would
improve peer-review and the editorial assessment of manuscripts, perhaps
with the aid of a checklist to standardise the criteria considered.

Similarly, published papers can be evaluated with a WoE approach, as was
done for the vitamin D and COVID-19 example. Papers such as Hastie et
al. (2020) provide nearly zero information for a vitamin D/COVID-19
association and can therefore be ignored, and certainly not included in
meta-analyses, where they can bias the results.

\subsection{Conclusion}\label{conclusion}

We have argued that in most biological and social science research the
probability of a parameter cannot equal the probability of a hypothesis.
First, qualitative background and contextual information is always
needed to interpret the parameter value in light of the hypothesis.
Second, multiple parameters associated with multiple outcomes that test
the same hypothesis are rarely equal and may even conflict. Third, the
value of a parameter (and its corresponding p-value or Bayes factor)
often depends on unknown and unmeasured characteristics of the
population or experiment as well as the statistical model.

Hence, statements such as ``the treatment is effective (\(p < 0.05\))''
makes little sense. Expanded, this statement means ``we conclude that
the treatment is effective \emph{because} \(p < 0.05\)'', but this is
\emph{non sequitur} and potentially misleading. We can only conclude
that ``the plausibility of the treatment being effective has increased
because the data are unlikely, given the null value of a parameter in a
statistical model.'' The extent to which the plausibility has increased
will depend on the study's ability to detect true effects (power) and to
avoid false positives, which can be quantified and represented as a WoE.

WoE provides a structured approach, but there is a risk that its
subjectivity and flexibility could be exploited to justify biased
interpretations, particularly in contentious or high-stakes research
areas. For instance, senior scientists entrenched in the dominant
paradigm could use the WoE approach to dismiss novel and promising
research avenues. The WoE method, however, promotes a more comprehensive
and transparent evaluation of evidence, which enables critical scrutiny
of both data and underlying assumptions, ultimately leading to more
robust and credible conclusions.

\section*{References}\label{references}
\addcontentsline{toc}{section}{References}

\phantomsection\label{refs}

\end{document}